\documentclass[submission,copyright,creativecommons]{eptcs}

\providecommand{\event}{Ninth International Symposium on Games,\\ Automata, Logics, and Formal Verification (GandALF'18)} % Name of the event you are submitting to
\usepackage{underscore}           % Only needed if you use pdflatex.

\usepackage[utf8]{inputenc}
\usepackage[english]{babel}
\usepackage{amsmath}
\usepackage{amsfonts}
\usepackage{amssymb}
\usepackage{amsthm}
\usepackage{tikz}
\usetikzlibrary{shapes}
\usetikzlibrary{automata}

\usetikzlibrary{shapes.multipart}
\usetikzlibrary{decorations.pathmorphing}
\usepackage{caption}

\def\rest#1#2{#1_{\restriction#2}}
\def\outcome#1#2{\langle #1 \rangle_{#2}}

\DeclareMathOperator{\Succ}{Succ}
\DeclareMathOperator{\Min}{min}
\DeclareMathOperator{\Max}{max}
\DeclareMathOperator{\Occ}{Occ}
\DeclareMathOperator{\Inf}{Inf}

\DeclareMathOperator{\Hist}{Hist}
\DeclareMathOperator{\Plays}{Plays}

\DeclareMathOperator{\First}{First}

\DeclareMathOperator{\Gain}{Gain}
\DeclareMathOperator{\Win}{Obj}

\newcommand{\init}{init}
\newcommand{\W}{\Omega}

\DeclareMathOperator{\rP}{\mathbf{P}}
\DeclareMathOperator{\Wit}{\mathcal{P}}

\usepackage[colorinlistoftodos,prependcaption,textsize=tiny]{todonotes}
\usepackage{xargs}                      % Use more than one optional parameter in a new commands

\newcommandx{\unsure}[2][1=]{\todo[linecolor=red,backgroundcolor=red!25,bordercolor=red,#1]{#2}}
\newcommandx{\change}[2][1=]{\todo[linecolor=blue,backgroundcolor=blue!25,bordercolor=blue,#1]{#2}}
\newcommandx{\info}[2][1=]{\todo[linecolor=BlueGreen,backgroundcolor=OliveGreen!25,bordercolor=OliveGreen,#1]{#2}}
\newcommandx{\improvement}[2][1=]{\todo[linecolor=SpringGreen,backgroundcolor=SpringGreen!25,bordercolor=SpringGreen,#1]{#2}}
\newcommandx{\thiswillnotshow}[2][1=]{\todo[disable,#1]{#2}}

\theoremstyle{definition}
\newtheorem{defi}{Definition}
\newtheorem{example}{Example}

\theoremstyle{theorem}
\newtheorem{thm}{Theorem}
\newtheorem{prop}{Proposition}
\newtheorem{lemma}{Lemma}
\newtheorem{corollary}{Corollary}

\theoremstyle{remark}

\makeatletter
\let\c@defi\c@thm
\let\c@lemma\c@thm
\let\c@prop\c@thm

\let\c@corollary\c@thm
\let\c@remark\c@thm
\let\c@example\c@thm
\let\c@proposition\c@thm
\makeatother

\author{Thomas Brihaye
\institute{Université de Mons (UMONS), Belgium}
\email{thomas.brihaye@umons.ac.be}
\and
Véronique Bruyère
\institute{Université de Mons (UMONS), Belgium}
\email{veronique.bruyere@umons.ac.be}
\and
Aline Goeminne
\institute{Université de Mons (UMONS), Belgium \\
Université libre de Bruxelles (ULB), Belgium }
\email{aline.goeminne@umons.ac.be}
\and
Jean-François Raskin
\institute{Université libre de Bruxelles (ULB), Belgium }
\email{jraskin@ulb.ac.be}
}
\def\titlerunning{\quad Constrained Existence Problem for Weak SPE with $\omega$-Regular Boolean Objectives}
\def\authorrunning{T. Brihaye, V. Bruyère, A. Goeminne \& J.-F. Raskin}

\title{Constrained Existence Problem for Weak Subgame Perfect Equilibria with $\omega$-Regular Boolean Objectives\thanks{The four authors are supported by COST Action GAMENET CA 16228 and by the FNRS PDR project ``Subgame perfection in graph games'' (T.0088.18). Jean-Fran\c cois Raskin is supported by the ERC Starting Grant inVEST (279499), by the ARC project ``Non-Zero Sum Game Graphs: Applications to Reactive Synthesis and Beyond'' (F\'ed\'eration Wallonie-Bruxelles), and by the EOS project ``Verifying Learning Artificial Intelligence Systems'' (FNRS-FWO), and he is Professeur Francqui de Recherche funded by the Francqui foundation.}}
\begin{document}
\maketitle

\begin{abstract}
We study multiplayer turn-based games played on a finite directed graph such that each player aims at satisfying an $\omega$-regular Boolean objective. Instead of the well-known notions of Nash equilibrium (NE) and subgame perfect equilibrium (SPE), we focus on the recent notion of weak subgame perfect equilibrium (weak SPE), a refinement of SPE. In this setting, players who deviate can only use the subclass of strategies that differ from the original one on a finite number of histories. We are interested in the constrained existence problem for weak SPEs. We provide a complete characterization of the computational complexity of this problem: it is P-complete for Explicit Muller objectives, NP-complete for Co-B\"uchi, Parity, Muller, Rabin, and Streett objectives, and PSPACE-complete for Reachability and Safety objectives (we only prove NP-membership for B\"uchi objectives). We also show that the constrained existence problem is fixed parameter tractable and is polynomial when the number of players is fixed. All these results are based on a fine analysis of a fixpoint algorithm that computes the set of possible payoff profiles underlying weak SPEs.
\end{abstract}

%!TEX root=main.tex
\section{Introduction}

\emph{Two-player zero-sum graph games} with $\omega$-regular objectives are the classical mathematical model to formalize the reactive synthesis problem~\cite{PnueliR89,Thomas95}. More recently, generalization from zero-sum to non zero-sum, and from two players to $n$ players have been considered in the literature, see e.g.~\cite{berwanger07,BrenguierCHPRRS16,BRS-concur15,BRS14,BMR14,KHJ06,FismanKL10,KupfermanPV14,Ummels06} and the surveys~\cite{Bruyere17,GU08}. 
Those extensions are motivated by two main limitations of the classical setting. First, zero-sum games assume a fully antagonistic environment while this is often not the case in practice: the environment usually has its own goal.  While the fully antagonistic assumption is simple and sound (a winning strategy against an antagonistic environment is winning against any environment that pursues its own objective), it may fail to find a winning
strategy even if solutions exist when the objective of the environment is accounted.
Second, modern reactive systems are often composed of several modules, and each module has its own specification and should be considered as a player on its own right. This is why we need to consider \emph{$n$-player graph games}.

For $n$-player graph games, solution concepts like \emph{Nash equilibria} (NEs)~\cite{nash50} are natural to consider. 
A strategy profile is an NE if no player has an incentive to deviate unilaterally from his strategy, \emph{i.e.} no player can strictly improve on the outcome of the strategy profile by changing his strategy only.
In the context of sequential games (such as games played on graphs), NEs allow for non-credible threats that rational players should not carry out. To avoid non-credible threats, refinements such as \emph{subgame perfect equilibria} (SPEs)~\cite{osbornebook} have been advocated. A strategy profile is an SPE if it is an NE in all the subgames of the original game. So players need to play rationally in all subgames, and this ensures that non-credible threats cannot exist. For applications of this concept to $n$-player graph games, we refer the reader to~\cite{BrihayeBDG12,KHJ06,Ummels06}. 

In~\cite{BrihayeBMR15}, the notion of {\em weak} subgame perfect equilibrium (weak SPE) is introduced, and it is shown how it can be used to study the existence SPEs (possibly with contraints) in quantitative reachability games.  While an SPE must be resistant to any unilateral deviation of one player, a weak SPE must be resistant to deviations restricted to deviating strategies that differ from the original one on a \emph{finite number} of histories only. In~\cite{Bruyere0PR17} the authors study general conditions on the structure of the game graph and on the preference relations of the players that guarantee the existence of a weak SPE for quantitative games. Weak SPEs retain most of the important properties of SPEs and they coincide with them when the payoff function of each player is continuous (see e.g.~\cite{fudenberg1991game}). Weak SPEs are also easier to characterize and to manipulate algorithmically. We refer the interested reader to~\cite{BrihayeBMR15,Bruyere0PR17} for further justifications of their interest, as well as for related work on NEs and SPEs.

\paragraph{Main contributions}
In this paper, we concentrate on graph games with \emph{$\omega$-regular Boolean objectives}. While SPEs, and thus weak SPEs, are always guaranteed to exist in such games, we here study the computational complexity of the {\em constrained existence problem} for weak SPEs, \emph{i.e.} equilibria in which some designated players have to win and some other ones have to loose. More precisely, our main results are as follows:
\begin{itemize}
  	\item We study the constrained existence problem for games with Reachability, Safety, B\"uchi, Co-B\"uchi, Parity, Explicit Muller, Muller, Rabin, and Streett objectives. We provide a \emph{complete characterization} of the computational complexity of this problem for all the classes of objectives with one exception: B\"uchi objectives. The problem is P-complete for Explicit Muller objectives, it is NP-complete for Co-B\"uchi, Parity, Muller, Rabin, and Streett objectives, and it is PSPACE-complete for Reachability and Safety objectives. In case of B\"uchi objectives, we show membership to NP but we fail to prove hardness. 
	\item Our complexity results rely on the identification of a \emph{symbolic witness} for the constrained existence of a weak SPE, the size of which allows us to prove NP/PSPACE-membership. As the constrained existence problem is PSPACE-complete for Reachability and Safety objectives, symbolic witnesses as compact as those for the other objectives cannot exist unless NP $=$ PSPACE. The identification of symbolic witnesses is obtained thanks to a \emph{fixpoint algorithm} that computes the set of all possible payoff profiles underlying weak SPEs.
	\item When the number of players is fixed, we show that the constrained existence problem can be solved in \emph{polynomial} time for all $\omega$-regular objectives. We also prove that it is \emph{fixed parameter tractable} where the parameter is the number of players, for Reachability, Safety, B\"uchi, Co-B\"uchi, and Parity objectives. For Rabin, Streett, and Muller objectives, we still establish fixed parameter tractability but we need to consider some additional parameters depending on the objectives. These tractability results are obtained by a fine analysis of the complexity of the fixpoint algorithm mentioned previously. 
\end{itemize}

\paragraph{Related work and additional contributions} 
In~\cite{GU08,Ummels06}, a tree automata-based algorithm is given to decide the constrained existence problem for SPEs on graph games with $\omega$-regular objectives defined by parity conditions. A complexity gap is left open: this algorithm executes in EXPTIME and NP-hardness of the decision problem is proved. In this paper, we focus on weak SPEs for which we provide precise complexity results for the constrained existence problem. We also observe that our results on Reachabilty and Safety objectives transfer from weak SPEs to SPEs: the constrained existence problem for SPEs is PSPACE-complete for those objectives. {\em Quantitative} Reachability objectives are investigated in~\cite{BrihayeBMR15} where it is proved that the constrained existence problem for weak SPEs and SPEs is decidable, but its exact complexity is left open.

In~\cite{BrihayeBMR15,Bruyere0PR17,FleschKMSSV10}, the existence of (weak) SPEs in graph games is established using a construction based on a fixpoint.  Our fixpoint algorithm is mainly inspired by the fixpoint technique of~\cite{Bruyere0PR17}. However, we provide complexity results based on this fixpoint while transfinite induction is used in~\cite{Bruyere0PR17}. Furthermore, we have modified the technique of~\cite{Bruyere0PR17} in a way to get a fixpoint that contains exactly all the possible payoff profiles of weak SPEs. This is necessary to get a decision algorithm for the constrained existence problem. 
 
Profiles of strategies with finite-memory are more appealing from a practical point of view. It is shown in~\cite{Ummels06} that when there exists an SPE in a graph game with $\omega$-regular objectives, then there exists one that uses finite-memory strategies and has the same payoff profile. Thanks to the symbolic witnesses, we have refined those results for weak SPEs.

\paragraph{Structure of the paper}  In Section~\ref{sec:prelim}, we recall the notions of $n$-player graph games and of (weak) SPE, and we state the studied constrained existence problem. In Section~\ref{section:charac}, we provide a fixpoint algorithm that computes all the possible payoff profiles for weak SPEs on a given graph game. From this fixpoint, we derive symbolic witnesses of weak SPEs. In Section~\ref{sec:classes}, we study the complexity classes of the constrained existence problem for all objectives. We also prove the fixed parameter tractability of the constrained existence problem and we show that it is in polynomial time when the number of players is fixed. In Section~\ref{sec:conc}, we give a conclusion and propose future work.

%!TEX root=main.tex
\section{Preliminaries}
\label{sec:prelim}

In this section, we introduce multiplayer graph games in which each player aims to achieve his Boolean objective. We focus on classical $\omega$-regular objectives, like Reachability, B\"uchi, aso. We recall two classical concepts of equilibria: Nash equilibrium and subgame perfect equilibrium~(see \cite{GU08}). We also recall weak variants of these equilibria as proposed in~\cite{BrihayeBMR15,Bruyere0PR17}. We conclude the section by the constrained existence problem that is studied in this paper.

%\subsection{Multiplayer Boolean games}
\paragraph{Multiplayer Boolean games} A \emph{multiplayer Boolean game} is a tuple $\mathcal{G} = (\Pi, V, (V_i)_{i \in \Pi}, E, (\Gain_i)_{i \in \Pi})$ where $\Pi = \{ 1,2, \ldots,n \}$ is a finite set of $n$ \emph{players}, $G = (V,E)$ is a finite directed graph and for all $v \in V$ there exists $v' \in V$ such that $(v,v') \in E$, $(V_i)_{i \in \Pi}$ is a partition of $V$ between the players, and $\Gain = (\Gain_i)_{i \in \Pi}$ is a tuple of functions $\Gain_i: V ^\omega \rightarrow  \{0,1\}$ that assigns a Boolean value  to each infinite path of $G$ for player $i$.

A \emph{play} in $\mathcal{G}$ is an infinite  sequence of vertices $\rho = \rho_0 \rho_1 \ldots$ such that for all $k \in \mathbb{N}$, $(\rho_k, \rho_{k+1}) \in E$.  A \emph{history} is a finite sequence $h = h_0h_1 \ldots h_n$ ($n \in\mathbb{N}$) defined similarly. We denote the set of plays by $\Plays$ and the set of histories by $\Hist$. Moreover, the set $\Hist_i$ is the set of histories such that the last vertex $v$ is a vertex of player $i$, i.e. $v \in V_i$. The \emph{length} $|h|$ of $h$ is the number $n$ of its edges. A play $\rho$ is called a \emph{lasso} if it is of the form $\rho = h\ell^\omega$ with $h\ell \in \Hist$. Notice that $\ell$ is not necessary a simple cycle. The \emph{length of a lasso} $h\ell^\omega$ is the length of $h\ell$. For all $h\in \Hist$, we denote by $\First(h)$ the first vertex $h_0$ of $h$. We use notation $h < \rho$ when a history $h$ is prefix of a play (or a history)  $\rho$. Given a play $\rho = \rho_0\rho_1 \ldots$, the set $\Occ(\rho) = \{ v \in V \mid \exists k, \rho_k = v \}$ is the set of vertices \emph{visited} by $\rho$, and $\Inf(\rho) = \{ v \in V \mid \forall k, \exists j \geq k, \rho_j = v \}$ is the set of vertices \emph{infinitely often visited} by $\rho$. Given a vertex $v \in V$, $\Succ(v) = \{v' \mid (v, v') \in E \}$ is the set of successors of $v$, and $\Succ^*(v)$ is the set of vertices reachable from $v$ in $G$.

When an \emph{initial} vertex $v_0\in V$ is fixed, we call $(\mathcal{G}, v_0)$ an \emph{initialized game}. A play (resp. a history) of $(\mathcal{G},v_0)$ is a play (resp. a history) of $\mathcal{G}$ starting in $v_0$. The set of such plays (resp. histories) is denoted by $\Plays(v_0)$ (resp. $\Hist(v_0)$). We also use notation $\Hist_i(v_0)$ when these histories end in a vertex $v \in V_i$.

The goal of each player $i$ is to achieve his objective, \emph{i.e.}, to maximize his gain.

\begin{defi}[Objective]
	\label{defi:winningCondition}
	For each player $i \in \Pi$, let $\Win_i \subseteq V^\omega$ be his \emph{objective}. In the setting of multiplayer Boolean game, the gain function $\Gain_i$ is defined such that $\Gain_i(\rho) = 1$ (resp. $\Gain_i(\rho) = 0$) if and only if $\rho \in \Win_i$ (resp. $\rho \not \in \Win_i$).
	\end{defi}
	
An objective $\Win_i$ (or the related gain function $\Gain_i$) is \emph{prefix-independent} if for all $h \in V^*$ and $ \rho \in V^{\omega}$, we have $\rho \in \Win_i$ if and only if $h\rho \in \Win_i$. In this paper, we focus on classical \emph{$\omega$-regular} objectives: Reachability,  Safety, Büchi, Co-Büchi, Parity, Explicit Muller, Muller, Rabin, and Streett and we suppose that each player has the \emph{same type} of objective. For instance, we say that $\mathcal{G}$ is a \emph{Boolean game with B\"uchi objectives} to express that all players have a B\"uchi objective. 

\begin{defi}[Classical $\omega$-regular objective]
The set $\Win_i$ is a \emph{Reachability, Safety, Büchi, Co-Büchi, Parity, Explicit Muller, Muller, Rabin}, or \emph{Streett} objective for player~$i$ if and only if $\Win_i$ is composed of the plays $\rho$ satisfying:
\begin{itemize}
	\item \emph{Reachability}:  given $F \subseteq V$, $\Occ(\rho) \cap F \neq \emptyset$;
	\item \emph{Safety}: given $F \subseteq V$, $ \Occ(\rho) \cap F = \emptyset$;
	\item \emph{Büchi}: given $F \subseteq V$, $ \Inf(\rho) \cap F \neq \emptyset $;
	\item \emph{Co-Büchi}: given $F \subseteq V$, $ \Inf(\rho) \cap F = \emptyset$;
	\item \emph{Parity}: given a coloring function $\Omega : V \rightarrow \{1,\ldots,d\}$, $ \Max(\Inf(\Omega(\rho)))\footnote{Where $\Omega(\rho) = \Omega(\rho_0)\Omega(\rho_1)\ldots \Omega(\rho_n)\ldots$.} \text{ is even}$;
	\item \emph{Explicit Muller}: given $\mathcal{F}  \subseteq 2^V$, $\Inf(\rho)\in \mathcal{F}$; 
	\item \emph{Muller}: given a coloring function $\Omega : V \rightarrow \{1,\ldots,d\}$, and $\mathcal{F} \subseteq 2^{\Omega(V)}$, $\Inf(\Omega(\rho)) \in \mathcal{F}$;
	\item \emph{Rabin}: given $(G_j,R_j)_{1\leq j \leq k}$ a family of pair of sets $G_j,R_j \subseteq V$,\\there exists $j \in {1,\ldots,k}$ such that $\Inf(\rho) \cap G_j \neq \emptyset$ and $\Inf(\rho) \cap R_j = \emptyset$;
	\item \emph{Streett}: given $(G_j,R_j)_{1\leq j \leq k}$ a family of pair of sets $G_j,R_j \subseteq V$,\\ for all $j \in {1,\ldots,k}$,  $\Inf(\rho) \cap G_j = \emptyset$ or $\Inf(\rho) \cap R_j \neq \emptyset$.
\end{itemize}
\end{defi}

\noindent
All these objectives are prefix-independent except Reachability and Safety objectives.

A \emph{strategy} of a player $i\in \Pi$ is a function $\sigma_i: \Hist_i \rightarrow V$. This function assigns to each history $hv$ with $v \in V_i$, a vertex $v'$ such that $(v,v') \in E$. In an initialized game $(\mathcal{G},v_0)$, $\sigma_i$ needs only to be defined for histories starting in $v_0$. A play $\rho=\rho_0\rho_1\ldots$ is \emph{consistent} with  $\sigma_i$ if for all $\rho_k \in V_i$ we have that $\sigma_i(\rho_0 \ldots \rho_k) = \rho_{k+1}$. A strategy $\sigma_i$ is \emph{positional} if it only depends on the last vertex of the history, \emph{i.e.}, $\sigma_i(hv) = \sigma_i(v)$ for all $hv \in \Hist_i$. It is \emph{finite-memory} if it can be encoded by a deterministic \emph{Moore machine} ${\cal M} = (M, m_0, \alpha_u, \alpha_n)$ where $M$ is a finite set of states (the memory of the strategy), $m_0 \in M$ is the initial memory state, $\alpha_u\colon M \times V \rightarrow M$ is the update function, and $\alpha_n\colon M \times V_i \rightarrow V$ is the next-action function. The Moore machine $\cal M$ defines a strategy $\sigma_i$ such that $\sigma_i(h v) = \alpha_n(\widehat{\alpha}_u(m_0,h),v)$ for all histories $h v \in \Hist_i$, where $\widehat{\alpha}_u$ extends $\alpha_u$ to histories as expected. The \emph{size} of the strategy $\sigma_i$ is the size $|M|$ of its machine $\cal M$. Note that $\sigma_i$ is positional  when $|M| = 1$.

A \emph{strategy profile} is a tuple $\sigma = (\sigma_i)_{i\in \Pi}$ of strategies, one for each player. It is called positional (resp. finite-memory) if for all $i \in \Pi$, $\sigma_i$ is positional (resp. finite-memory).  Given an initialized game $(\mathcal{G}, v_0)$ and a strategy profile $\sigma$, there exists an unique play from $v_0$ consistent with each strategy $\sigma_i$. We call this play the \emph{outcome} of $\sigma$ and it is denoted by $\outcome{\sigma}{v_0}$. Let $p = (p_i)_{i \in \Pi} \in \{0,1\}^{|\Pi|}$, we say that $\sigma$ is a strategy profile \emph{with payoff} $p$ or that $\outcome{\sigma}{v_0}$ \emph{has payoff} $p$ if $p_i = \Gain_i(\outcome{\sigma}{v_0})$ for all $i \in \Pi$.

%\subsection{Solution concepts}
%\label{section:solutionConcepts}
\paragraph{Solution concepts}
In the multiplayer game setting, the solution concepts usually studied are \emph{equilibria} (see~\cite{GU08}). We here recall the concepts of Nash equilibrium and subgame perfect equilibrium, as well as some variants. We begin with the notion of deviating strategy.

Let $\sigma = (\sigma_i)_{i\in \Pi}$ be a strategy profile in an initialized Boolean game $(\mathcal{G},v_0)$. Given $i \in \Pi$, a strategy $\sigma'_i \neq \sigma_i$ is a \emph{deviating} strategy of player~$i$, and $(\sigma'_i, \sigma_{-i})$ denotes the strategy profile $\sigma$ where $\sigma'_i$ replaces $\sigma_i$. Such a strategy is a \emph{profitable deviation} for player~$i$ if $\Gain_i(\outcome{\sigma}{v_0}) < \Gain_i(\outcome{\sigma'_i, \sigma_{-i}}{v_0})$. We say that $\sigma'_i$ is  \emph{finitely deviating} from $\sigma_i$ if  $\sigma'_i$ and $\sigma_i$ only differ on a finite number of histories, and that $\sigma'_i$ is \emph{one-shot deviating} from $\sigma_i$ if $\sigma'_i$ and $\sigma_i$ only differ on $v_0$~\cite{BrihayeBMR15,Bruyere0PR17}.  

The notion of Nash equilibrium (NE) is classical: a strategy profile $\sigma$ in an initialized game $(\mathcal{G},v_0)$ is a \emph{Nash equilibrium} if no player has an incentive to deviate unilaterally from his strategy since he has no profitable deviation, \emph{i.e.}, for each $i \in \Pi$ and each deviating strategy $\sigma'_i$ of player $i$ from $\sigma_i$, the following inequality holds: $\Gain_i(\outcome{\sigma}{v_0}) \geq \Gain_i(\outcome{\sigma'_i, \sigma_{-i}}{v_0})$. In this paper we focus on two variants of NE:  weak/very weak NE~\cite{BrihayeBMR15,Bruyere0PR17}. 

\begin{defi}[Weak/very weak Nash equilibrium]
	A strategy profile $\sigma$ is a \emph{weak NE} (resp. \emph{very weak NE}) in $(\mathcal{G},v_0)$ if, for each player $i\in \Pi$, for each finitely deviating (resp. one-shot) strategy $\sigma'_i$ of player~$i$ from $\sigma_i$, we have $\Gain_i(\outcome{\sigma}{v_0}) \geq \Gain_i(\outcome{\sigma'_i, \sigma_{-i}}{v_0})$.
\end{defi}

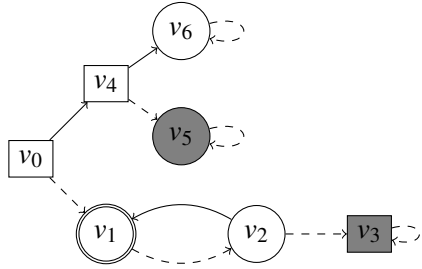
\begin{figure}
	\centering
	\scalebox{1}{

	\begin{tikzpicture}[]
		\node[draw] (v0) at (0,0){$v_0$};
		\node[draw,accepting,circle] (v1) at (1,-1){$v_1$};
		\node[draw,circle] (v2) at (3,-1){$v_2$};
		\node[draw, fill=gray] (v3) at (4.5,-1){$v_3$};
		\node[draw] (v4) at (1,1){$v_4$};
		\node[draw,circle, fill=gray] (v5) at (2,0.3){$v_5$};
		\node[draw,circle] (v6) at (2,1.7){$v_6$};

		\draw[->,dashed] (v0) to (v1);
		\draw[->,dashed] (v1) to [bend right] (v2);
		\draw[->] (v2) to [bend right] (v1);

		\draw[->,dashed] (v2) to (v3);
		\draw[->] (v0) to  (v4);

		\draw[->,dashed] (v3) to [loop right] (v3);
		\draw[->,dashed] (v4) to (v5);
		\draw[->] (v4) to (v6);
		\draw[->,dashed] (v5) to [loop right] (v5);
		\draw[->,dashed] (v6) to [loop right] (v6);

	\end{tikzpicture}}
	\caption{Example of a Boolean game with B\"uchi objectives}
	\label{fig:weakNEvsNE}   
\end{figure}

\begin{example}
\label{ex:weakNE}
Figure~\ref{fig:weakNEvsNE} illustrates an initialized Boolean game $(\mathcal{G},v_0)$ with B\"uchi objectives  in which there exists a weak NE that is not an NE. In this game, player~1 (resp. player~2) owns round (resp. square) vertices and wants to visits $v_1$ (resp. $v_3$ or $v_5$) infinitely often. The positional strategy profile $\sigma = (\sigma_1,\sigma_2)$ is depicted by dashed arrows, its outcome is equal to $\outcome{\sigma}{v_0}=v_0v_1v_2v_3^\omega$, and $\sigma$ has payoff $(0,1)$. Notice that player 1 has an incentive to deviate from his strategy $\sigma_1$ with a strategy $\sigma'_1$ that goes to $v_1$ for all histories ending in $v_2$. This is indeed a profitable deviation for him since $\Gain(\outcome{(\sigma'_1, \sigma_2)}{v_0}) = (1,0)$. So, $\sigma$ is not an NE. Nevertheless, it is a weak NE because $\sigma'_1$ is the only profitable deviation and it is not finitely deviating (it differs from $\sigma_1$ on all histories of the form $v_0(v_1v_2)^k$ for $k \in \mathbb{N}\backslash \{0 \}$).
\end{example} 

When considering games played on graphs, a well-known refinement of NE is the concept of \emph{subgame perfect equilibrium} (SPE) which a strategy profile being an NE in each subgame. Variants of weak/very weak SPE can also be studied as done with NEs. Formally, given an initialized Boolean game $(\mathcal{G},v_0)$ and a history $hv \in \Hist(v_0)$, the initialized game $(\rest{\mathcal{G}}{h},v)$ is called a \emph{subgame}\footnote{Notice that $(\mathcal{G},v_0)$ is subgame of itself.} of $(\mathcal{G},v_0)$ such that $\rest{\mathcal{G}}{h} = (\Pi, V, (V_i)_{i\in \Pi}, E, \rest{\Gain}{h})$ and $\Gain_{i\restriction h}(\rho) = \Gain_i(h\rho)$ for all $i \in \Pi$ and $\rho \in V^{\omega}$. Moreover if $\sigma_i$ is a strategy for player~$i$ in $(\mathcal{G},v_0)$, then $\sigma_{i\restriction h}$ denotes the strategy in $(\rest{\mathcal{G}}{h},v)$ such that for all histories $h'\in \Hist_i(v)$, $\sigma_{i\restriction h}(h') = \sigma_i(hh')$. Similarly, from a strategy profile $\sigma$ in $(\mathcal{G},v_0)$, we derive the strategy profile $\rest{\sigma}{h}$ in $(\rest{\mathcal{G}}{h},v)$. The play $\outcome{\rest{\sigma}{h}}{v}$ is called a \emph{subgame outcome} of $\sigma$. 

\begin{defi}[Subgame perfect equilibrium and weak/very weak subgame perfect equilibrium]
	A strategy profile $\sigma$ is a \emph{(resp. weak, very weak) subgame perfect equilibrium} in $(\mathcal{G},v_0)$ if for all $hv \in \Hist(v_0)$, $\rest{\sigma}{h}$ is a (resp. weak, very weak) NE in $(\rest{\mathcal{G}}{h},v)$.
\end{defi}

When one needs to show that a strategy profile is a weak SPE, the next proposition is very useful because it states that it is enough to consider one-shot deviating strategies.  
\begin{prop}[\cite{BrihayeBMR15}] \label{prop:equivWSPEetVWSPE}
	A strategy profile $\sigma$ is a weak SPE if and only if $\sigma$ is a very weak SPE.
\end{prop}

\begin{example}
\label{ex:weakSPE}
In Example~\ref{ex:weakNE} is given a weak NE $\sigma$ in the game $(\mathcal{G},v_0)$ depicted in Figure~\ref{fig:weakNEvsNE}. This strategy profile is also a very weak SPE (and thus a weak SPE by Proposition~\ref{prop:equivWSPEetVWSPE}). For instance, in the subgame $(\rest{\mathcal{G}}{h},v)$ with $h = v_0v_1$ and $v = v_2$, the only one-shot deviating strategy $\sigma'_1$ is such that $\sigma'_1$ coincides with $\sigma_{1\restriction h}$ except that $\sigma'_1(v_2) = v_1$. This is not a profitable deviation for player~$1$ in $(\rest{\mathcal{G}}{h},v)$. Notice that $\sigma$ is not an SPE since it is not an NE as explained in Example~\ref{ex:weakNE}.
\end{example}

%In general, the notions of SPE and weak SPE are not equivalent (see Example~\ref{ex:weakSPE}). Nevertheless they coincide for the class of Boolean games with Reachability objectives. 
%
%
%\begin{prop} 
%\label{prop:reach}
%Let $\sigma$ be a strategy profile in an initialized Boolean game $(\mathcal{G},v_0)$ with Reachability objectives. Then $\sigma$ is an SPE if and only if $\sigma$ is a weak SPE.
%\end{prop}

%\subsection{Constraint problem}
\paragraph{Constraint problem}
It is proved in~\cite{Bruyere0PR17} that there always exists a weak SPE in Boolean games.  In this paper, we are interested in solving the following \emph{constraint problem}:

\begin{defi}[Constraint problem]
\label{decidabilityProblem}
Given $(\mathcal{G},v_0)$ an initialized Boolean game and thresholds $x,y \in \{0,1\}^{|\Pi|}$, decide whether there exists a weak SPE in $(\mathcal{G},v_0)$ with payoff $p$  such that $x \leq p \leq y$.\footnote{The order $\leq$ is the componentwise order, that is, $x_i \leq p_i \leq y_i$, for all $i \in \Pi$.}
\end{defi}

In the next sections, we solve the constraint problem for the classical $\omega$-regular objectives. The complexity classes that we obtain are shown in Table~\ref{tab:complexity}; they are detailed in Section~\ref{sec:classes}. The case of B\"uchi objectives remains open, since we only propose a non-deterministic algorithm in polynomial time but no matching lower bound. In Section~\ref{sec:classes}, we also prove that the constraint problem for weak SPEs is fixed parameter tractable and becomes polynomial when the number of players is fixed. All these results are based on a characterization of the set of possible payoffs of a weak SPE, that is described in Section~\ref{section:charac}. 

\begin{table}[h!]
\centering
\caption{Complexity classes of the constraint problem for $\omega$-regular objectives}
\scalebox{0.9}{
\begin{tabular}{|l|c|c|c|c|c|c|c|c|}
\hline
                & Explicit Muller & Co-Büchi  & Parity  & Muller & Rabin & Streett & Reachability & Safety  \\ \hline\hline
P-complete       & $\times$  &   &   &       &   &   &   &        \\ \hline
NP-complete    &   		& $\times$ & $\times$ &  $\times$  & $\times$      &  $\times$     &     &  \\ \hline
PSPACE-complete      &   &  &  &        &   &    & $\times$  & $\times$\\ \hline
\end{tabular}}
\label{tab:complexity}
\end{table}

%!TEX root=main.tex
\section{Characterization}
\label{section:charac}

In this section our aim is twofold: first, we characterize the set of possible payoffs of weak SPEs and second, we show how it is possible to build a weak SPE given a set of lassoes with some ``good properties". Those characterizations work for Boolean games with \emph{prefix-independent gain functions}. We make this hypothesis all along Section~\ref{section:charac}.

%\subsection{Remove-Adjust procedure}
%\label{sec:RemoveAdjust}
\paragraph{Remove-Adjust procedure}
Let $(\mathcal{G},v_0)$ be an initialized Boolean game with prefix-independent gain functions. The computation of the set of all the payoffs of weak SPEs in $(\mathcal{G},v_0)$ is inspired by a fixpoint procedure explained in~\cite{Bruyere0PR17}. Each vertex $v$ is \emph{labeled} by a set of payoffs $p \in \{0,1\}^{|\Pi|}$.
Initially, these payoffs are those for which there exists a play in $\Plays(v)$ with payoff $p$. Then step by step, some payoffs are removed for the labeling of $v$ as soon as we are sure they cannot be the payoff of $\rest{\sigma}{h}$ in a subgame $(\rest{\mathcal{G}}{h},v)$ for some weak SPE $\sigma$. (Notice that the value of $h$ is not important since the gain functions are prefix independent. This is why we only focus on $v$ and not on $hv$.) 
When a fixpoint is reached, the labeling of the initial vertex $v_0$ exactly contains all the payoffs of weak SPEs in $(\mathcal{G},v_0)$. 

We formally proceed as follows. For all $v \in V$, we define the initial labeling of $v$ as:
$$\rP_0(v) = \{ p \in \{0,1\}^{|\Pi|} \mid \text{ there exists } \rho \in \Plays(v) \text{ such that } \Gain(\rho) = p \}.$$

Then for each step $k \in \mathbb{N} \setminus \{0\}$, we compute the set $\rP_k(v)$ by alternating between two operations: \emph{Remove} and \emph{Adjust}. To this end, we need to introduce the notion of $(p,k)$-labeled play. Let $p$ be a payoff and $k$ be a step, a play $\rho= \rho_0\rho_1\rho_2 \ldots$ is \emph{$(p,k)$-labeled} if for all $j \in \mathbb{N}$ we have $p \in \rP_k(\rho_j)$, that is, $\rho$ visits only vertices that are labeled by $p$ at step $k$. We first give the definition of the Remove-Adjust procedure and then give some intuition about it.

\begin{defi}[Remove-Adjust procedure] 
\label{def:remove}

Let $k \in \mathbb{N} \setminus \{0\}$.
\begin{itemize}
\item If $k$ is odd, process the \emph{Remove} operation: 
\begin{itemize}
\item If for some $v \in V_i$ there exists $p \in \rP_{k-1}(v)$ and $v' \in \Succ(v)$ such that $p_i < p'_i$ for all $p' \in \rP_{k-1}(v')$,
then $\rP_k(v) = \rP_{k-1}(v) \backslash \{p\}$ and for all $u \neq v$, $\rP_k(u) = \rP_{k-1}(u)$. 
\item If such a vertex $v$ does not exist, then $\rP_k(u) = \rP_{k-1}(u)$ for all $u \in V$.
\end{itemize}
\item If $k$ is even, process the \emph{Adjust} operation: 
\begin{itemize}
\item If some payoff $p$ was removed from $\rP_{k-2}(v)$ (that is, $\rP_{k-1}(v) = \rP_{k-2}(v)\setminus \{p\}$), then 
\begin{itemize}
\item For all $u \in V$ such that $p \in \rP_{k-1}(u)$, check whether there still exists a $(p,k-1)$-labeled play with payoff $p$ from $u$. If it is the case, then $\rP_{k}(u) = \rP_{k-1}(u)$, otherwise $\rP_{k}(u) = \rP_{k-1}(u) \setminus \{p \}$.
\item For all $u \in V$ such that $p \notin \rP_{k-1}(u)$:
$\rP_{k}(u) = \rP_{k-1}(u).$
\end{itemize}
\item Otherwise $\rP_k(u) = \rP_{k-1}(u)$ for all $u \in V$.
\end{itemize}
\end{itemize}
\end{defi}

Let us explain the Remove operation. Let $p$ that labels vertex $v$. This means that it is the payoff of a potential subgame outcome of a weak SPE that starts in $v$. Suppose that $v$ is a vertex of player~$i$ and $v$ has a successor $v'$ such that $p_i < p'_i$ for all $p'$ labeling $v'$. Then $p$ cannot be the payoff of $\rest{\sigma}{h}$ in the subgame $(\rest{\mathcal{G}}{h},v)$ for some weak SPE $\sigma$ and some history $h$, otherwise player~$i$ would have a profitable (one-shot) deviation by moving from $v$ to $v'$ in this subgame.

Now it may happen that for another vertex $u$ having $p$ in its labeling, all potential subgame outcomes of a weak SPE from $u$ with payoff $p$ necessarily visit vertex $v$. As $p$ has been removed from the labeling of $v$, these potential plays do no longer survive and $p$ is also removed from the labeling of $u$ by the Adjust operation.

We can state the existence of a fixpoint of the sequences $(\rP_k(v))_{k\in \mathbb{N}}$, $v \in V$, in the following meaning:
 
 \begin{prop}[Existence of a fixpoint]
	\label{prop:existenceFixpoint}
 There exists an even natural number $k^* \in \mathbb{N}$ such that for all $v \in V$, $\rP_{k^*}(v) = \rP_{k^*+1}(v) = \rP_{k^*+2}(v)$. 
 \end{prop} 

\begin{example}
	\label{exemple:pointFixe}
	We illustrate the different steps of the Remove-Adjust procedure on the example depicted in Figure~\ref{fig:weakNEvsNE}, and we display the result of this computation in Table~\ref{example:remove-adjustProc}. Initially, the sets $\rP_0(v)$, $v \in V$, contains all payoffs $p$ such that there exists a play $\rho \in \Plays(v)$ with $\Gain(\rho) = p$. 
	At step $k = 1$, we apply a Remove operation to $v = v_4$ (this is the only possible $v$): $v$ is a vertex of player~$i = 2$ and $v$ has a successor $v' = v_5$ such that $(0,1) \in \rP_0(v_5)$. Therefore $(0,0)$ is removed from $\rP_0(v_4)$ to get $\rP_1(v_4)$. By definition of the Remove operation, the other sets $\rP_0(u)$ are not modified and are thus equal to $\rP_1(u)$.
 	At step $k=2$, we apply an Adjust operation. The only way to obtain the payoff $(0,0)$ from $v_0$ is by visiting $v_4$ with the play $v_0v_4v_6^\omega$. As there does not exist a $((0,0),1)$-labeled play with payoff $(0,0)$ anymore, we have to remove $(0,0)$ from $\rP_1(v_0)$. The other sets $\rP_1(v)$ remain unchanged. 
	At step $k=3$, the Remove operation removes payoff $(1,0)$ from $\rP_2(v_0)$ due to the unique payoff $(0,1)$ in $\rP_2(v_4)$.
	At step $k=4$, the Adjust operation leaves all sets $\rP_3(v)$ unchanged. 
	Finally at step $k=5$, the Remove operation also leaves all sets $\rP_4(v)$ unchanged, and the fixpoint is reached.
	
	\begin{table}[h!]
		\caption{Computation of the fixpoint on the example of Figure~\ref{fig:weakNEvsNE}}
		\label{example:remove-adjustProc}
		\centering
		\scalebox{0.8}{
		\begin{tabular}{|l||l|l|l|l|l|l|l|}
			\hline
			& $v_0$ & $v_1$ & $v_2$ & $v_3$ & $v_4$ & $v_5$ & $v_6$\\
			\hline
			\hline
			$\rP_0(v)$ & $\{ (0,0),(1,0),(0,1) \}$ & $\{(1,0),(0,1)\}$ &$\{(1,0),(0,1)\}$ &$\{(0,1)\}$ & $\{\mathbf{(0,0)},(0,1)\}$&$\{(0,1)\}$& $\{(0,0)\}$\\
			\hline
			$\rP_1(v)$ & $\{ \textbf{(0,0)},(1,0),(0,1) \}$ & $\{(1,0),(0,1)\}$ &$\{(1,0),(0,1)\}$ &$\{(0,1)\}$ & $\{(0,1)\}$&$\{(0,1)\}$& $\{(0,0)\}$\\
			\hline 
			$\rP_2(v)$ & $\{\textbf{(1,0)},(0,1) \}$ & $\{(1,0),(0,1)\}$ &$\{(1,0),(0,1)\}$ &$\{(0,1)\}$ & $\{(0,1)\}$&$\{(0,1)\}$& $\{(0,0)\}$\\
			\hline
			$\rP_3(v)$  & $\{(0,1) \}$ & $\{(1,0),(0,1)\}$ &$\{(1,0),(0,1)\}$ &$\{(0,1)\}$ & $\{(0,1)\}$&$\{(0,1)\}$& $\{(0,0)\}$\\
			\hline
			$\rP_4(v) = \rP_{k^*}(v)$  & $\{(0,1) \}$ & $\{(1,0),(0,1)\}$ &$\{(1,0),(0,1)\}$ &$\{(0,1)\}$ & $\{(0,1)\}$&$\{(0,1)\}$& $\{(0,0)\}$\\
			\hline
		\end{tabular}}
	\end{table}
\end{example}

%\subsection{Characterization and good symbolic witness}
%\label{sec: witness}
\paragraph{Characterization and good symbolic witness}
The fixpoint $\rP_{k^*}(v)$, $v \in V$, provides a characterization of the payoffs of all weak SPEs as described in the following theorem. This result is in the spirit of the classical \emph{Folk
    Theorem} which characterizes the payoffs of all NEs in infinitely
  repeated games (see for instance~\cite[Chapter~8]{fudenberg1991game}).

\begin{thm}[Characterization]
 \label{folkThm}
 Let $(\mathcal{G},v_0)$ be an initialized Boolean game with prefix-independent gain functions. Then there exists a weak SPE $\sigma$ with payoff $p$ in $(\mathcal{G},v_0)$ if and only if $\rP_{k^*}(v) \neq \emptyset$ for all $v \in \Succ^*(v_0)$ and $p \in \rP_{k^*}(v_0)$.%\footnote{We use notation $p_0 \in \{0,1\}^{|\Pi|}$ to highlight that this is the payoff of $\sigma$ from vertex $v_0$. It should not be confused with any component $p_i$, $i \in \Pi$, of a payoff $p$.}
 \end{thm}

In this theorem, only sets $\rP_{k^*}(v)$ with $v \in \Succ^*(v_0)$ are considered. Indeed subgames $(\rest{\mathcal{G}}{h},v)$ of $(\mathcal{G},v_0)$ deals with histories $hv \in \Hist(v_0)$, that is, with vertices $v$ reachable from $v_0$. %The rest of this section is devoted to the proof of Theorem~\ref{folkThm}. 
%
%We begin with a lemma that states that if a payoff $p$ survives at step $k$ in the labeling of $v$, this means that there exists a play with payoff $p$ from $v$ that only visits vertices also labeled by $p$.
%
%\begin{lemma}
% \label{intermediaryResults:lemmaLabeledPlay}
% For all even $k$ and in particular for $k = k^*$, $p$ belongs to $\rP_k(v)$ if and only if there exists a $(p,k)$-labeled play $\rho \in \Plays(v)$ such that $\Gain(\rho) = p$.
% \end{lemma}
% 
The proof of Theorem~\ref{folkThm} uses the concept of (good) symbolic witness defined hereafter. Some intuition about it is given after the definitions.

\begin{defi}[Symbolic witness] 
	\label{def:symbolicWitnesses}
	 Let $(\mathcal{G},v_0)$ be an initialized Boolean game with prefix-independent gain functions. Let $I \subseteq (\Pi \cup \{0\}) \times V$ be the set 
	 $$I = \{(0,v_0)\} ~\cup~ \{ (i,v') \mid \text{ there exists } (v,v')\in E \text{ such that } v, v' \in \Succ^*(v_0) \text{ and } v \in V_i \}.$$
	 A  \emph{symbolic witness} is a set $\Wit = \{\rho_{i,v} \mid (i,v) \in I \}$ such that each $\rho_{i,v} \in \Wit$ is a lasso of $G$ with $\First(\rho_{i,v}) = v$ and with length bounded by $2 \cdot |V|^2$.
\end{defi}

A symbolic witness has thus at most $|V| \cdot |\Pi| + 1$ lassoes (by definition of $I$) with polynomial length.

\begin{defi}[Good symbolic witness]
	\label{def:Good}
	A symbolic witness $\Wit$ is \emph{good} if for all $\rho_{j,u}, \,\, \rho_{i,v'} \in \Wit$, for all vertices $v \in \rho_{j,u}$ such that $v \in V_i$ and $v' \in \Succ(v)$, we have 
	 $\Gain_i(\rho_{j,u}) \geq \Gain_i(\rho_{i,v'})$.
\end{defi}

\begin{figure}[h!]
	\centering
	\begin{tikzpicture}
		\node[draw,circle,minimum width=20pt](u) at (0,0){$u$};
		\node(point1) at (1.5,0){$\ldots$};
		\node[draw,circle,minimum width=20pt](v) at (3,0){$v$};
		\node(vi) at (3,-0.7){$\in V_i$};
		\node[draw,circle,minimum width=20pt](v') at (3.5,1){$ v'$};
		\node(point2) at (5,1){$\ldots$};
		\node[draw,circle,minimum width=20pt](blanc1) at (6.5,1){}; 
		\node(point3) at (8,1){$\ldots$};
		\node[draw,circle,minimum width=20pt](blanc2) at (9.5,1){}; 
		\node (path1) at (10.5,1){$\rho_{i,v'}$};
		\node(point4) at (5,0){$\ldots$};
		\node[draw,circle,minimum width=20pt](blanc3) at (6.5,0){}; 
		\node (point5) at (8,0){$\ldots$};
		\node[draw,circle,minimum width=20pt](blanc4) at (9.5,0){};
		\node (path2) at (10.5,0){$\rho_{j,u}$};	
		
		\draw[->] (u) -- (1,0);
		\draw[->] (2,0) -- (v);
		\draw[->] (v) -- (v');
		\draw[->] (v') -- (4.5,1);
		\draw[->] (5.5,1) --(blanc1);
		\draw[->] (blanc1) -- (7.5,1);
		\draw[->] (8.5,1) -- (blanc2);
		\draw[->] (blanc2) to [bend right] (blanc1);
		
		\draw[->] (v) -- (4.5,0);
		\draw[->] (5.5,0) -- (blanc3);
		\draw [->] (blanc3) -- (7.5,0);
		\draw[->] (8.5,0) -- (blanc4);
		\draw[->] (blanc4) to [bend left] (blanc3);
	\end{tikzpicture}
	\caption{The condition of Definition~\ref{def:Good}}
	\label{fig:Good}
\end{figure}
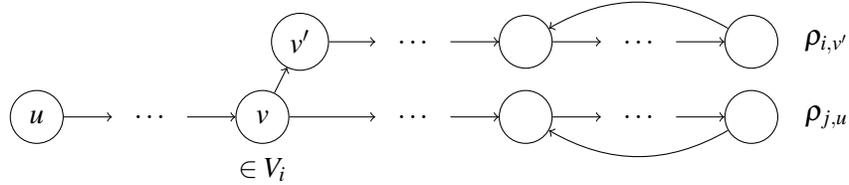

The condition of Definition~\ref{def:Good} is depicted in Figure~\ref{fig:Good}.

Let us now give some intuition. A strategy profile $\sigma$ in $(\mathcal{G},v_0)$ induces an infinite number of subgame outcomes $\outcome{\rest{\sigma}{h}}{v}$, $hv \in \Hist(v_0)$. A symbolic witness $\Wit$ is a \emph{compact} representation of $\sigma$. It is a finite set of lassoes that represent some subgame outcomes of $\sigma$: the lasso $\rho_{0,v_0}$ of $\Wit$ represents the outcome $\outcome{\sigma}{v_0}$, and its other lassoes $\rho_{i,v'}$ represents the subgame outcome $\outcome{\rest{\sigma}{h}}{v'}$ for some particular histories $hv' \in \Hist(v_0)$. The index $i$ records that player~$i$ can move from $v$ (the last vertex of $h$) to $v'$ (with the convention that $i = 0$ for the outcome $\outcome{\sigma}{v_0}$). When $\sigma$ is a weak SPE, the related symbolic witness $\Wit$ is good, that is, its lassoes avoid profitable one-shot deviations between them.

\begin{example} 
We come back to our running example. The weak SPE of Example~\ref{ex:weakSPE} depicted in Figure~\ref{fig:weakNEvsNE} has payoff $p = (0,1)$. A symbolic witness $\Wit$ of $\sigma$ is given in Table~\ref{table:setOfSymbolicWitnesses} which is here composed of all the subgame outcomes of $\sigma$. One can check that $\Wit$ is a good symbolic witness. For instance, consider its lassoes $\rho_{0,v_0} = v_0v_1v_2v_3^\omega$ and $\rho_{1,v_1} = v_1v_2v_3^\omega$, the vertex $v_2 \in V_1$ of $\rho_{0,v_0}$ and the edge $(v_2,v_1)$. We have $\Gain_1(\rho_{0,v_0}) \geq \Gain_1(\rho_{1,v_1})$. Indeed in the subgame $(\rest{\mathcal{G}}{v_0v_1},v_2)$, player~$1$ has no profitable one-shot deviation by using the edge $(v_2,v_1)$.
\begin{table}[h!]
	\centering
	\caption{An example of good symbolic witness}
	\begin{tabular}{|l|l|l|l|l|l|l|l|l|l|}
		\hline
		$(i,v)$ & $(0,v_0)$ &$(2,v_4)$ & $(1,v_2)$ & $(1,v_1)$ & $(1,v_3)$ & $(2,v_5)$ & $(2,v_6)$ & $(1,v_5)$ & $(1,v_6)$ \\
		\hline
		lasso & $v_0v_1v_2v_3^\omega$ & $v_4v_5^\omega$ & $v_2v_3^\omega$ & $v_1v_2v_3^\omega$ &  $v_3^\omega$ &  $v_5^\omega$  & $v_6^\omega$ & $v_5^\omega$ & $v_6^\omega$ \\
		\hline
		payoff & $(0,1)$ & $(0,1)$ & $(0,1)$ & $(0,1)$ & $(0,1)$& $(0,1)$ & $(0,0)$ & $(0,1)$ & $(0,0)$\\
		\hline
	\end{tabular}
		\label{table:setOfSymbolicWitnesses}
\end{table}
\end{example}

%In Proposition~\ref{prop:lasseFolkThm} below, we are going to prove that there exists a weak SPE if and only if there exists a good symbolic witness, and that the existence of this witness is equivalent to the non-emptiness of the fixpoint $\rP_{k^*}(v)$, $v \in V$. In this way, we will prove Theorem~\ref{folkThm}. We will see that the lassoes $\rho_{i,v}$ of a good symbolic witness can be constructed from $(p,k^*$)-labeled plays for well-chosen payoffs $p \in \rP_{k^*}(v)$.
Theorem~\ref{folkThm} is a direct consequence of the next proposition.

\begin{prop}
	\label{prop:lasseFolkThm}
	 Let $(\mathcal{G},v_0)$ be an initialized Boolean game with prefix-independent gain functions. The following assertions are equivalent:
	\begin{enumerate}
		\item There exists a weak SPE with payoff $p$ in $(\mathcal{G},v_0)$; \label{Lasse:1}
		\item $\rP_{k^*}(v) \neq \emptyset$ for all $v \in \Succ^*(v_0)$ and $p \in \rP_{k^*}(v_0);$ \label{Lasse:2}
		\item There exists a good symbolic witness $\Wit$ that contains a lasso $\rho_{0,v_0}$ with payoff $p$;  \label{assert:existenceWitnesses}
		\item There exists a finite-memory weak SPE $\sigma$ with payoff $p$ in $(\mathcal{G},v_0)$ such that the size of each strategy $\sigma_i$ is in $\mathcal{O}(|V|^3 \cdot |\Pi|)$. \label{assert:existenceWSPEpoly}
	\end{enumerate}	
\end{prop}

Let us give a sketch of proof. For $(1 \Rightarrow 2)$, if there exists a weak SPE $\sigma$, then for all $hv \in \Hist(v_0)$ and all $k$, the payoff $\Gain(\outcome{\rest{\sigma}{h}}{v})$ survives in $\rP_k(v)$ after both a Remove and an Adjust operation. This shows that the fixpoint is non empty. For $(2 \Rightarrow 3)$, if each $\rP_{k^*}(v)$ contains some payoff $p$, then there exists a $(p,k^*)$-labeled play $\rho \in \Plays(v)$ such that $\Gain(\rho) = p$, and this play can be supposed to be a lasso of size at most $2 \cdot |V|^2$ by \cite[Proposition 3.1]{BouyerBMU15}. We then construct a symbolic witness $\Wit$ composed of some of those lassoes: $\rho_{0,v_0} \in \Wit$ is a lasso extracted from $\rP_{k^*}(v_0)$, and each other lasso $\rho_{i,v'} \in \Wit$ is extracted from $\rP_{k^*}(v')$ such that its payoff $p'$ satisfies $p'_i = \Min\{ q_i \mid q \in \rP_{k^*}(v') \}$. The way the fixpoint is constructed guarantees that if $p \in \rP_{k^*}(v)$ with $v \in V_i$, we have $p_i \geq p'_i$ (by the Remove operation). The latter property implies that $\Wit$ is good. For $(3 \Rightarrow 4)$, from the existence of a good symbolic witness $\Wit$, it is possible to construct a strategy profile $\sigma$ such that its subgame outcomes are the lassoes of $\Wit$, and that is a weak SPE because $\Wit$ is good. Moreover the strategies of $\sigma$ are finite-memory with size in $\mathcal{O}(|V|^3 \cdot |\Pi|)$ because $\sigma$ is constructed from at most $|V| \cdot |\Pi| + 1$ lassoes with length bounded by $2 \cdot |V|^2$. Finally $(4 \Rightarrow 1)$ is immediate.

Recall that Proposition~\ref{prop:lasseFolkThm} only works for prefix-independent gain functions. Nevertheless a Boolean game with Reachability (resp. Safety) objectives can be transformed into a Boolean game with B\"uchi (resp. Co-B\"uchi) objectives.  The construction is classical: it stores inside the vertices the set of players~$i$ who have already visited $F_i$, where $F_i$ is the set they want to visit (resp. to avoid). Consequently we have the next result from equivalence $(1 \Leftrightarrow 4)$ of Proposition~\ref{prop:lasseFolkThm}.
 
\begin{corollary}
\label{cor:finitememory}
Let $(\mathcal{G},v_0)$ be an initialized Boolean game. There exists a weak SPE in $(\mathcal{G},v_0)$ if and only if there exists a finite-memory weak SPE $\sigma$ with the same payoff. Moreover, the size of each strategy $\sigma_i$ is 
\begin{itemize}
\item in $\mathcal{O}(|V|^3 \cdot |\Pi|)$ for B\"uchi, Co-B\"uchi, Parity, Explicit Muller, Muller, Rabin, and Streett objectives,
\item in $\mathcal{O}(|V|^3 \cdot |\Pi| \cdot 2^{3\cdot |\Pi|} )$ for Reachability and Safety objectives.
\end{itemize}

\end{corollary}

%!TEX root=main.tex
\section{Complexity of the constraint problem}
\label{sec:classes}

%In this section, we study the complexity classes of the constraint problem for Boolean games with classical $\omega$-regular objectives, except the case of Explicit Muller objectives that is postponed to Section~\ref{sec:ExplicitMuller} (see Table~\ref{tab:complexity}). The concept of good symbolic witness is essential in this study. 

In this section, we first study the complexity classes of the constraint problem for Boolean games with all classical $\omega$-regular objectives (except Explicit Muller objectives). The concept of good symbolic witness is essential in this study. Second, we show that the constraint problem is $P$-complete for Explicit Muller objectives, that it is fixed parameter tractable for the other classical $\omega$-regular objectives, and that it becomes polynomial when the number of players is fixed. Those results rely on the fixpoint algorithm given in the previous section.

%!TEX root=main.tex
%\subsection{NP-completeness}
%\label{sec:NP}

%We first prove that the constraint problem for co-Büchi, Parity, Muller, Rabin, and Streett objectives is NP-complete, and that it is in NP for B\"uchi objectives. 

%\begin{thm}
%\label{thm:NP}
%The constraint problem for Boolean games with co-Büchi, Parity, Muller, Rabin, and Streett objectives is NP-complete. It is in NP for B\"uchi objectives. 
%\end{thm}

\paragraph{Complexity classes of the constraint problem} The exact complexities of the constraint problem are stated in the following theorem.

\begin{thm}
\label{thm:NPetPSPACE} The constraint problem for Boolean games
\begin{itemize}
\item is PSPACE-complete for Reachability and Safety objectives;
\item is NP-complete for co-Büchi, Parity, Muller, Rabin, and Streett objectives; it is in NP for B\"uchi objectives. 
\end{itemize}
\end{thm}

Let us give a sketch of proof. For Part 2 of Theorem~\ref{thm:NPetPSPACE}, as the considered objectives are prefix-independent, we can apply Proposition~\ref{prop:lasseFolkThm}, in particular equivalence $(1 \Leftrightarrow 3)$. To prove NP-membership, given thresholds $x, y \in \{0,1\}^{|\Pi|}$, a nondeterministic polynomial algorithm works as follows: guess a set $\Wit$ composed of at most $|\Pi| \cdot |V| + 1$ lassoes of length bounded by $2 \cdot |V|^2$ and check that $\Wit$ is a good symbolic witness that contains a lasso $\rho_{0,v_0}$ with payoff $p$ such that $x \leq p \leq y$. The NP-hardness is obtained thanks to a polynomial reduction from SAT, using exactly the same reduction as proposed in~\cite{Ummels08} for the constraint problem for NEs in Boolean games with co-Büchi objectives. Notice that this reduction does not work for B\"uchi objectives. For Part 1 of Theorem~\ref{thm:NPetPSPACE}, we transform a given Boolean game with Reachability (resp. Safety) objectives into a Boolean game with B\"uchi (resp. Co-B\"uchi) objectives (as explained at the end of Section~\ref{section:charac}). With this transformation, a good symbolic witness has an exponential number of lassoes however still with polynomial size. To prove PSPACE-membership, we here consider an alternating Turing machine such that along each execution, the players $\vee$ and $\wedge$ play in a turned based way during a polynomial number of turns: player~$\vee$ proposes lassoes (taken in a good symbolic witness $\Wit$ if it exists) and player~$\wedge$ tries to show that $\Wit$ is not good (by proposing a deviation between two lassoes that do not respect Definition~\ref{def:Good}). This Turing machine works in polynomial time, and as PSPACE $=$ APTIME, we get PSPACE-membership. The PSPACE-hardness is obtained with a polynomial reduction from QBF. This reduction is more involved than the one used to prove NP-hardness. Indeed the reduction for NP-hardness already works for NEs whereas the reduction for PSPACE-hardness really exploits the subgame perfect aspects. 

Notice that the constraint problem for SPEs (instead of weak SPEs) for Boolean games with Reachability and Safety objectives is also PSPACE-complete. Indeed one can prove that weak SPEs and SPEs are equivalent notions for Reachability objectives. The case of Safety objectives needs another approach, it results from the proof of Theorem~\ref{thm:NPetPSPACE}. 
%Recall that weak SPEs and SPEs are equivalent notions for Reachability objectives (Proposition~\ref{prop:reach}). It follows from Theorem~\ref{thm:PSPACE} that the constraint problem for SPEs (instead of weak SPEs) for Boolean games with Reachability objectives is PSPACE-complete. We will see later (in Section~\ref{sec:cor}, from the proof of Theorem~\ref{thm:PSPACE}) that the constraint problem for SPEs is also PSPACE-complete for Safety objectives. 

\begin{corollary}
\label{cor:PSPACE}
The constraint problem for SPEs in Boolean games with Reachability and Safety objectives is PSPACE-complete.
\end{corollary}

%\input {PSPACEcompleteness}

%!TEX root=main.tex
%\section{Fixed parameter tractability and Explicit Muller objectives}
%\label{sec:FPT}

\paragraph{Fixed parameter tractability and Explicit Muller objectives} We now study the complexity of the constraint problem for Explicit Muller objectives, and its fixed parameter tractability for the other objectives. We recall that a \emph{parameterized language} $L$ is a subset of $\Sigma^* \times \mathbb{N}$, where $\Sigma$ is a finite alphabet, the second component being the parameter of the language. It is called \emph{fixed parameter tractable} if there is an algorithm that determines whether $(x,t) \in L$ in $f(t) \cdot |x|^c$ time, where $c$ is a constant independent of the parameter $t$ and $f$ is a computable function depending on $t$ only. We also say that $L$ belongs to the class FPT. Intuitively, a language is in FPT if there is an algorithm running in polynomial time with respect to the input size times some computable function on the parameter. 
We refer the interested reader to \cite{DowneyF99} for more details on parameterized complexity.

\begin{thm}
	\label{thm:ExplicitMullerPcomplete}
	The constraint problem for multiplayer Boolean games with Explicit Muller objectives is P-complete.	
\end{thm}

\begin{thm} \label{thm:FPT}
Let $\mathcal{G}$ be a Boolean game.
\begin{enumerate}
\item The constraint problem is in FPT for Reachability, Safety, B\"uchi, co-B\"uchi, Parity, Muller, Rabin, and Streett objectives. The parameters are
\begin{itemize}
\item the number $|\Pi|$ of players for Reachability, Safety, B\"uchi, co-B\"uchi, and Parity objectives,
\item the number $|\Pi|$ of players and the numbers $k_i$, $i \in \Pi$, of pairs $(G^i_j,R^i_j)_{1 \leq j \leq k_i}$, for Rabin and Streett objectives, and
\item the number $|\Pi|$ of players, the numbers $d_i$, $i \in \Pi$, of colors and the sizes $|\mathcal{F}_i|$, $i \in \Pi$, of the families of subsets of colors for Muller objectives.
\end{itemize}
\item When the number $|\Pi|$ of players is fixed, for all these kinds of objectives, the constraint problem can be solved in polynomial time.
\end {enumerate}
\end{thm}

Notice that in Theorem~\ref{thm:FPT}, to obtain fixed parameter tractability for Rabin, Streett, and Muller objectives, in addition to the number of players, we also have to consider the parameter equal to the size of the objective description. Nevertheless, when the number of players is fixed, we get polynomial time complexity for all types of objectives.  

The results of Theorems~\ref{thm:ExplicitMullerPcomplete} and~\ref{thm:FPT} do not rely on the concept of good symbolic witness (as for Theorem~\ref{thm:NPetPSPACE}) but rather on the following algorithm based on Theorem~\ref{folkThm} and called the \emph{decision algorithm}. Given a Boolean game $(\mathcal{G},v_0)$ and thresholds $x, y \in \{0,1\}^{|\Pi|}$, 
\begin{itemize}
\item Compute the initial sets $\rP_0(v)$, $v \in V$, and repeat the Remove-Adjust procedure (see Definition~\ref{def:remove}) until reaching the fixpoint $\rP_{k^*}(v)$, $v \in V$, 
\item Then check whether $\rP_{k^*}(v) \neq \emptyset$ for all $v \in \Succ^*(v_0)$ and whether there exists a payoff $p \in \rP_{k^*}(v_0)$ such that $x \leq p \leq y$.
\end{itemize}

%We call this algorithm the \emph{decision algorithm} and its first part computing the fixpoint the \emph{fixpoint algorithm}.

%!TEX root=main.tex
%\subsection{Complexity of the decision algorithm}
%\label{section:complexityFixpoint}

\paragraph{Time complexity of the decision algorithm} Let us study the time complexity of the decision algorithm in terms of three parameters: $(i)$ $\mathcal{O}(\init)$: the complexity of computing $\rP_0(v)$ for some given vertex $v$, $(ii)$ $m = \Max_{v\in V} |\rP_0(v)|$: the maximum number of payoffs in the sets $\rP_0(v)$, $v \in V$, and $(iii)$ $\mathcal{O}(path)$: the complexity of determining whether there exists a play with a given payoff $p$ from a given vertex $v$. (This test is required in both the computation of $\rP_0(v)$ and the Adjust operation.)

\begin{lemma} \label{lem:compl}
The time complexity of the decision algorithm is in $\mathcal{O}(m^3 \cdot |V| \cdot |\Pi| \cdot init \cdot path \cdot (|V| + |E|))$.
\end{lemma}

Expressing the complexity in this way is useful in the proof of both Theorem~\ref{thm:ExplicitMullerPcomplete} and~\ref{thm:FPT}. We do not claim that the given complexity is the tightest one but this is enough for our purpose.

Let us focus on parameter $\mathcal{O}(path)$. We have the next lemma. 
%The purpose of this section is to prove the next lemma stating the complexity $\mathcal{O}(path)$ for all kinds of $\omega$-regular objectives.  

\begin{lemma} \label{lem:FPT}  
Let $\mathcal{G}$ be a Boolean game. Let $p \in \{0,1\}^{|\Pi|}$ and $v \in V$.
\begin{enumerate}
\item Determining whether there exists a play with payoff $p$ from $v$ is
\begin{itemize}
\item in polynomial time for B\"uchi, co-B\"uchi, Explicit Muller, and Parity objectives,
\item in $\mathcal{O}(2^{|\Pi|} (|V| + |E|))$ time for Reachability and Safety objectives, and
\item in ${\cal O}(2^L \cdot M + (L^L \cdot |V|)^5 )$ time for Rabin, Streett, and Muller objectives, where $L = 2^\ell$ and
\begin{itemize}
\item $\ell = \Sigma_{i = 1}^{|\Pi|} 2 \cdot k_i$ and $M = \mathcal{O}(\Sigma_{i = 1}^{|\Pi|} 2 \cdot k_i)$ such that for each player~$i \in \Pi$, $k_i$ is the number of his pairs $(G^i_j,R^i_j)_{1 \leq j \leq k_i}$ in the case of Rabin and Streett objectives, and
\item $\ell = \Sigma_{i = 1}^{|\Pi|} d_i$ and $M = \mathcal{O}(\Sigma_{i = 1}^{|\Pi|} |\mathcal{F}_i| \cdot d_i)$ such that for each player~$i \in \Pi$, $d_i$ (resp. $|\mathcal{F}_i|$), is the number of his colors (the size of his family of subsets of colors) in the case of Muller objectives.
\end{itemize}
\end{itemize}
\item When the number $|\Pi|$ of players is fixed, for all these kinds of objectives, the existence of a play with payoff $p$ from $v$ can be solved in polynomial time.
\end{enumerate}
\end{lemma}

The general approach to prove this lemma is the following one. A play with payoff $p$ from $v$ in a Boolean game $\mathcal{G}$ is a play satisfying an objective $\W$ equal to the conjunction of objectives $\Win_i$ (when $p_i = 1$) and of objectives $V^\omega \setminus \Win_i$ (when $p_i = 0$). It is nothing else than an infinite path in the underlying graph $G = (V,E)$ satisfying some particular $\omega$-regular objective $\W$. The existence of such paths is a well studied problem. For instance, if each $\Win_i$ is a Parity objective, as $V^\omega \setminus \Win_i$ is again a Parity objective, $\W$ is thus a conjunction of Parity objectives which is a Streett objective~\cite{ChatterjeeHP07}. Deciding the existence of a path satisfying a Streett objective can be done in polynomial time~\cite{EmersonL87}. 
Now for the other classes of objectives $\Win_i$, we use known results about \emph{two-player zero-sum games} $(G,\W)$, where player~$A$ aims at satisfying a certain objective $\W$ whereas player $B$ tries to prevent him to satisfy it.  A classical problem is to decide whether player~$A$ has a winning strategy that allows him to satisfy $\W$ against any strategy of player~$B$, see for instance~\cite{Bruyere17,2001automata}. When player~$A$ is the only one to play, the existence of a winning strategy for him is equivalent to the existence of a path satisfying $\W$ (see \cite[Section 3.1]{Bruyere17}). This is exactly the problem considered in Lemma~\ref{lem:FPT}. Notice that when each $\Win_i$ is a Rabin (resp. Streett, Muller) objective, the objective $\W$ is a Boolean combination of B\"uchi objectives. Deciding the existence of a winning strategy for player~$A$ in the game $(G,\W)$ when $\W$ is such an objective can be done in ${\cal O}(2^L \cdot M + (L^L \cdot |V|)^5 )$ time with $L = 2^\ell$, such that $M$ is the number of disjunctions and conjunctions in the Boolean combination of B\"uchi objectives and $\ell$ is the number of its B\"uchi objectives~\cite{BruyereHR17}. Finally, the proof of part 2 of Lemma~\ref{lem:FPT} requires a fine complexity analysis to get polynomial time complexity.

\medskip
Let us conclude this section with the sketches of proof for Theorems~\ref{thm:ExplicitMullerPcomplete} and~\ref{thm:FPT}. Theorem~\ref{thm:FPT} is easy to obtain from Lemmas~\ref{lem:compl} and~\ref{lem:FPT} and the following observations on parameters $m$ and $\mathcal{O}(\init)$. As each gain function $\Gain_i$ takes its values in $\{0,1\}$, $m$ is bounded by $2^{|\Pi|}$, and by definition of $\rP_0(v)$,  $\mathcal{O}(\init)$ is in $\mathcal{O}(2^{|\Pi|} \cdot path)$. For the proof of Theorem~\ref{thm:ExplicitMullerPcomplete}, to get P-membership, we also apply Lemmas~\ref{lem:compl} and~\ref{lem:FPT} but we need to show that $\mathcal{O}(m)$ and $\mathcal{O}(\init)$ are polynomial. This possible by showing that $m$ is linearly bounded by the sum of the sizes $|\mathcal{F}_i|$, where $\mathcal{F}_i$ it the Explicit Muller objective of each player~$i$.

\section{Conclusion and future work}
\label{sec:conc}
In this paper, we have studied the computational complexity of the constraint problem for weak SPEs. We were able to obtain precise complexities for all the classical classes of $\omega$-regular objectives (see Table~\ref{tab:complexity}), with one exception: we have proved NP-membership for B\"uchi objectives and failed to prove NP-hardness. We have also shown that the constraint problem can be solved in polynomial time when the number of players is fixed. Finally, we have provided some fixed parameter tractable algorithms when the number of players is considered as a parameter of the problem, for Reachability, Safety, B\"uchi, Co-B\"uchi, and Parity  objectives. For the other Rabin, Streett, and Muller objectives, we also had to consider the size of the objective description as a parameter to obtain fixed parameter tractability. In a future work, we want to understand if the use of this second parameter is really necessary.

By characterizing the exact complexity of the constraint problem for Reachability and Safety objectives, we have obtained that this problem for SPEs (as for weak SPEs) is PSPACE-complete for those objectives. In the future, we intend to investigate the complexity of the other classes of $\omega$-regular objectives for SPEs. It would be also interesting to extend the study to quantitative games. For instance the constraint problem for (weak) SPEs in reachability quantitative games is decidable~\cite{BrihayeBMR15} but its complexity is unknown.

\bibliographystyle{eptcs} 

\bibliography{biblio}

%\newpage
%\appendix

%\input{appendix}

%\input{circuit}
	
\end{document}